\documentstyle[amssymb,aps,epsfig,twocolumn]{revtex}

\begin{document}
\title{Cooper pairs as resonances}
\author{M. Fortes, M.A. Sol\'{\i }s}
\address{Instituto de F\'{\i }sica, Universidad Nacional Aut\'{o}noma de
M\'{e}xico \\
Apdo. Postal 20-364, 01000 M\'{e}xico, DF, Mexico}
\author{M. de Llano}
\address{Instituto de Investigaciones en Materiales, Universidad Nacional
Aut\'{o}noma de M\'{e}xico \\
Apdo. Postal 70-360, 04510 M\'{e}xico, DF, Mexico}
\author{V.V. Tolmachev*}
\address{Facultad de Ciencias \& Instituto de F\'{\i}sica, Universidad Nacional Aut\'onoma de M\'exico \\
Apdo. Postal 20-364, 01000 M\'{e}xico, DF, Mexico\\
{\small Received (January 15, 2001); (Revised March 5, 2001)}}

\maketitle
\noindent {\bf Abstract}
\begin{abstract}
Using the Bethe-Salpeter (BS) equation, Cooper pairing can be generalized to
include contributions from holes as well as particles from the ground state of
either an ideal Fermi gas (IFG) or of a BCS many-fermion state. The 
BCS model interfermion interaction is employed throughout.  In contrast to the better-known original Cooper pair problem for either two particles or two holes, the generalized Cooper equation 
in the IFG case has no real-energy solutions. Rather, it
possesses
two complex-conjugate solutions with purely imaginary energies. This
implies that
the IFG ground state is unstable when an attractive interaction is switched
on. However, solving the BS equation for the BCS ground state reveals two types 
of {\it real} solutions: one describing moving (i.e., having nonzero total, or center-of-mass, momenta) Cooper pairs as resonances (or bound composite particles with a {\it finite} lifetime), and another exhibiting superconducting collective
excitations analogous to Anderson-Bogoliubov-Higgs RPA modes. A Bose-Einstein-condensation-based picture of superconductivity is addressed. 
\end{abstract}


\smallskip
\noindent {\bf 1. Introduction}
\smallskip
\smallskip

The original Cooper pair (CP) equation \cite{Coop} is a two-electron
Schr\"{o}dinger equation in momentum representation with a given two-body
interaction (having some \textit{attraction}) but includes \textit{%
ad hoc} restrictions on the magnitudes of both electron wavevectors $%
{\mathbf{k}}_{1}$, ${\mathbf{k}}_{2}$, namely $ {k}_{1} > k_{_{F}}$, ${k}_{2} > k_{_{F}}$, where $k_{_F}$
is the electron Fermi wavenumber for an ideal nonrelativistic
many-electron system. One then seeks the energy eigenvalues of a
CP bound state and its corresponding wave-function.

Since there is no rigorous derivation of the original CP equation, several authors 
\cite{Gor,Gor2,Tolm,Tolm2} reformulated the complete CP problem without neglecting holes, using the mathematically \textit{exact} BS equation 
approach \cite{BS} applied to the system, in search of two-particle bound states in the presence of other system electrons. Such a treatment allows generalizing \cite{Tolma} several approaches in superconductivity theory, including the BCS, the BCS-Bose crossover, and the Bose-Einstein condensation pictures. But if the BS equation is based merely on the IFG ground state
one obtains purely imaginary solutions, suggesting that the ground state is
unstable in the presence of attractive interactions of some kind. This is an
{\it instability} of  the IFG ground state with respect to the creation of two-particle (2p-) or two-hole (2h-) resonant
states, a situation analogous to the classical problem of {\it
hydrodynamic instability} \cite{Betch}. \\

\smallskip
\noindent {\bf 2. BS equation based on BCS ground state}
\smallskip
\smallskip

Consider, however, the generalized
two-component, two-electron BS equation based {\it not} on the IFG
ground state \cite{Gor,Gor2,Tolm,Tolm2} but rather on
the BCS ground state.
We introduce the Bogoliubov-Valatin $u$, $v$ transformation of electron
Fermi operators $a_{{\mathbf{k}},\sigma }$ to new Fermi operators $\alpha _{{\mathbf{k}},\sigma }$, namely 
\begin{equation}
\ a_{{\mathbf{k}},\sigma }=u_{k}\,\alpha _{{\mathbf{k}},\sigma }+2\sigma
\,v_{k}\,\,\alpha _{-{\mathbf{k}},-\sigma }^{+}, 
\end{equation}

\noindent where $\sigma =\pm \frac{1}{2}$ is the spin projection for an
electron state. The coefficients $u_{k},v_{k}$ are real and depend on $k$.
The many-electron system hamiltonian is 
\noindent $H=H_{0}+H_{int}$ where
\[
H_{0} = {\sum}_{{\mathbf{k}},\,\sigma } \left(\epsilon _{k} -E_{F}\right) \,a_{{\mathbf{k,}}\sigma }^{+}a_{{\mathbf{k,}}\,\sigma }, 
\]
\vspace{0.6cm}
\[ 
\bigskip H_{int} =\frac{1}{2L^{3}}{\sum }^{\prime}_{{\mathbf{k}}_{1},{\mathbf{k}}_{2},%
{\mathbf{k}}_{1}^{^{\prime }},\,{\mathbf{k}}_{2}^{^{\prime }},\sigma _{1,}\sigma
_{2}} \nu(|{\mathbf{k}}_{1}-{\mathbf{k}}_{1}^{\prime}|)\times  
\]
\vspace{-0.7cm}
\begin{equation}
 \qquad \, \times a_{{\mathbf{k}}_{1}^{\prime }\sigma _{1}}^{+}a_{%
{\mathbf{k}}_{2}^{\prime }\sigma _{2}}^{+}a_{{\mathbf{k}}_{2}\sigma _{2}}a_{%
{\mathbf{k}}_{1}\sigma _{1}},  
\end{equation}
where $\epsilon _{{k}}=\hbar^{2}k^{2}/2m$ is the
kinetic energy of an electron of mass $m$; $E_{F}=\hbar
^{2}\,k_{F}^{2}/2m$ is the Fermi energy; $L$ is the system size; $\nu(q)$ the Fourier transform of the two-electron interaction potential; and the last sum is restricted by momentum conservation ${\mathbf{k}}%
_{1}^{\prime }+{\mathbf{k}}_{2}^{\prime }={\mathbf{k}}_{1}+{\mathbf{k}}_{2}$. This leads to the well-known BCS
hamiltonian \cite{Tolm,Tolm2,BCS}
\begin{equation}
H_{BCS} = U_{0} + {\sum }_{{\mathbf{k}},\,\sigma} E(k) \alpha_{\mathbf{k,}\,\sigma}^+ \alpha _{\mathbf{k,}\,\sigma},  
\label{BCS}
\end{equation}

\noindent with $U_{0}$ a generalized BCS ground-state energy

\[
U_{0}=2\sum_{\mathbf{k}}\left( \epsilon _{{k}}-E_{F}\right) v_{k}^{2}+
\]
\vspace{-0.5cm}
\[
 +L^{-3}\sum_{\mathbf{k,k}^{\prime }}\nu (\left| \mathbf{k-k}%
^{\prime }\right| )v_{k}u_{k}v_{k^{\prime }}u_{k^{\prime }}+ 
 \]
\vspace{-0.5cm}
\[
 +2L^{-3} \sum_{\mathbf{k,k}^{\prime }}\nu (0)v_{k}^{2}v_{k^{\prime
}}^{2}-L^{-3}\sum_{\mathbf{k,k}^{\prime }}\nu (\left| \mathbf{k-k}%
^{\prime }\right| )v_{k}^{2}v_{k^{\prime }}^{2}, 
\]

\noindent where
\[
 u_{k}^{2}=\frac{1}{2}\left[ 1+\frac{A(k)}{E(k)}\right] ,\quad v_{k}^{2}=%
\frac{1}{2}\left[ 1-\frac{A(k)}{E(k)}\right],
\]
\vspace{-0.5cm}
\begin{equation}
2u_{k}v_{k}=-\frac{B(k)}{E(k)}, \quad E(k)\equiv \sqrt{A^{2}(k)+B^{2}(k)}. 
\label{Ek}
\end{equation}
Here $B(k)$ plays the role of the original BCS energy gap $\Delta(k)$, and
\begin{eqnarray}
A(k) &\equiv&\epsilon _{\mathbf{k}} - E_{F}+  \nonumber \\
&+& 2L^{-3} \nu (0){\sum_{k\,^{^{\prime }}}} v_{k^{^{\prime
}}}^{2}- L^{-3} {\sum_{k\,^{^{\prime }}}}\nu (|{\mathbf{k}}-{\mathbf{k}}^{\prime}|)v_{k^{^{\prime }}}^{2},  \nonumber \\
B(k) &\equiv& L^{-3} {\sum_{k\,^{^{\prime }}} }\nu (|{\mathbf{k}}-{\mathbf{k}}^{\prime}|)u_{k^{^{\prime }}}v_{k^{^{\prime }}}.  
\end{eqnarray}

For the new interaction hamiltonian we obtain
\[
H_{int}^{\prime}\equiv H-H_{BCS}=
\]
\vspace{-0.7cm}
\[
 = \frac{1}{2L^{3}}{\sum }_{{\mathbf{k}}_{1}^{\prime },{\mathbf{k}}_{2}^{\prime
},{\mathbf{k}}_{1},{\mathbf{k}}_{2},\sigma _{1},\sigma _{2}}^{\prime }\nu (|%
{\mathbf{k}}_{1}-{\mathbf{k}}_{1}^{\prime }|)\times 
\] 
\vspace{-0.6cm}
\[\times L(k_{1},k_{1}^{\prime }) L(k_{2},k_{2}^{\prime }) \times 
\]
\vspace{-0.6cm}
\[ \times \alpha _{{\mathbf{k}}_{1}^{\prime }\sigma
_{1}}^{+}\alpha _{{\mathbf{k}}_{2}^{\prime }\sigma _{2}}^{+}\alpha _{{%
\mathbf{k}}_{2}\sigma _{2}}\alpha _{{\mathbf{k}}_{1}\sigma _{1}}+ 
\]
\vspace{-0.6cm}
\[+\frac{1}{4L^{3}}{\sum }_{{\mathbf{k}}_{1}^{\prime },{\mathbf{k}}%
_{2}^{\prime },{\mathbf{k}}_{1},{\mathbf{k}}_{2},\sigma _{1},\sigma
_{2}}^{\prime }\nu (|{\mathbf{k}}_{1}-{\mathbf{k}}_{1}^{\prime
}|)\times
\]
\vspace{-0.6cm}
\[ \times M(k_{1},k_{1}^{\prime }) M(k_{2},k_{2}^{\prime }) \times 
\]
 \vspace{-0.6cm}
\[ \times 2\sigma _{1}\alpha _{{\mathbf{k}}%
_{1}^{\prime }\sigma _{1}}^{+}\alpha _{-{\mathbf{k}}_{1},-\sigma
_{1}}^{+} 2\sigma _{2}\,\alpha _{-{\mathbf{k}}_{2}^{\prime },-\sigma
_{2}}\alpha _{{\mathbf{k}}_{2}\sigma _{2}}+ 
\]
\vspace{-0.6cm} 
\[ +\frac{1}{4L^{3}}{\sum }_{{\mathbf{k}}_{1}^{\prime },{\mathbf{k}}%
_{2}^{\prime },{\mathbf{k}}_{1},{\mathbf{k}}_{2},\sigma _{1},\sigma
_{2}}^{\prime} \nu (|{\mathbf{k}}_{1}-{\mathbf{k}}_{1}^{\prime
}|)\times 
\] 
\vspace{-0.6cm}
\[ \times M(k_{1},k_{1}^{\prime })L(k_{2},k_{2}^{\prime }) \times 
\]
 \vspace{-0.6cm}
\[ \times 2\sigma _{1}\alpha _{{\mathbf{k}}%
_{1}^{\prime }\sigma _{1}}^{+}\alpha _{-{\mathbf{k}}_{1},-\sigma
_{1}}^{+} \alpha _{{\mathbf{k}}_{2}^{\prime }\sigma _{2}}^{+}\alpha _{{%
\mathbf{k}}_{2}\sigma _{2}}+ 
\]
\vspace{-0.6cm} 
\[ +\frac{1}{4L^{3}}{\sum}^{\prime}_{{\mathbf{k}}_{1}^{\prime },{\mathbf{k}}%
_{2}^{\prime },{\mathbf{k}}_{1},{\mathbf{k}}_{2},\sigma _{1},\sigma _{2}} \nu (|{\mathbf{k}}_{1}-{\mathbf{k}}_{1}^{\prime
}|) \times
 \]
 \vspace{-0.6cm}
\[ \times L(k_{1},k_{1}^{\prime })M(k_{2},k_{2}^{\prime })\times 
\]
 \vspace{-0.6cm}
\[ \times \alpha _{{\mathbf{k}}_{1}^{\prime }\sigma
_{1}}^{+}\alpha _{{\mathbf{k}}_{1}\sigma _{1}} 2\sigma _{2}\,\alpha _{-{%
\mathbf{k}}_{2}^{\prime },-\sigma _{2}}\alpha _{{\mathbf{k}}_{2}\sigma _{2}}+ 
\]
\vspace{-0.6cm} 
\[ +\frac{1}{8L^{3}}{\sum}_{{\mathbf{k}}_{1}^{\prime },{\mathbf{k}}%
_{2}^{\prime },\mathbf{k}_{1},{\mathbf{k}}_{2},\sigma _{1},\sigma
_{2}}^{\prime }\nu (|{\mathbf{k}}_{1}-{\mathbf{k}}_{1}^{\prime
}|)\times 
\] 
\vspace{-0.6cm}
\[ \times M(k_{1},k_{1}^{\prime }) M(k_{2},k_{2}^{\prime })\times 
\]
 \vspace{-0.6cm} 
\[ \times 2\sigma _{1}\alpha _{{\mathbf{k}}%
_{1}^{\prime }\sigma _{1}}^{+}\alpha _{-{\mathbf{k}}_{1},-\sigma
_{1}}^{+} 2\sigma _{2}\,\alpha _{{\mathbf{k}}_{2}^{\prime }\sigma
_{2}}^{+}\alpha _{-{\mathbf{k}}_{2},-\sigma _{2}}^{+}+ 
\]
\vspace{-0.6cm} 
\[ +\frac{1}{8L^{3}}{\sum }_{{\mathbf{k}}_{1}^{\prime },{\mathbf{k}}%
_{2}^{\prime },\mathbf{k}_{1},{\mathbf{k}}_{2},\sigma _{1},\sigma
_{2}}^{\prime} \nu (|{\mathbf{k}}_{1}-{\mathbf{k}}_{1}^{\prime
}|)\times 
\] 
\vspace{-0.6cm}
\[ \times M(k_{1},k_{1}^{\prime }) M(k_{2},k_{2}^{\prime }) \times
\] 
\vspace{-0.6cm}
\[ \times 2\sigma _{1}\,\alpha _{-{\mathbf{k}}%
_{1}^{\prime },-\sigma _{1}}\alpha _{{\mathbf{k}}_{1}\sigma _{1}} 2\sigma
_{2}\,\alpha _{-{\mathbf{k}}_{2}^{\prime },-\sigma _{2}}\alpha _{{\mathbf{k}}%
_{2}\sigma _{2}}.
 \]

\noindent where $L(k,k^{\prime })\equiv u_{k}\,u_{k^{\prime }}-v_{k}\,v_{k^{\prime
}}\,$ and  $M(k,k^{\prime })\equiv u_{k}\,v_{k^{\prime }}+u_{k^{\prime
}}\,v_{k}\,.$ 

To obtain the BS equation  based on the BCS ground state consider the Feynman
diagrams of perturbation theory based on this ground state, where (\ref{BCS}) is the 
new unperturbed hamiltonian $H_{0}^{^{\prime }}$. We now have the usual arrowed
electron lines labeled by ${\mathbf{k}}$, $E$, $\sigma$ to which we associate the BCS
unperturbed Green's function 
\begin{equation}
{\mathcal{G}}_{0}({\mathbf{k}}, E, \sigma) = \frac{\hbar }{i}\frac{1}{%
-E + E(k)-i\varepsilon },  
\end{equation}

\noindent where $E(k)$ is given by (\ref{Ek}). There exist 
four-line-end double vertices of six different kinds
(see Fig. 1) where the interfermion interaction is denoted by dashed lines.
To a double vertex type (a) of Fig. 1, with two outgoing line ends with indices $({\mathbf{k}}_1^{\prime}$, ${{E}_1^{\prime}}$, $\sigma _{1})$ and $({\mathbf{k}}_2^{\prime}$, ${{E}_2^{\prime}}$, $\sigma _{2})$ along  with two incoming line ends with indices $({\mathbf{k}}_1$, ${{E}_1}$, $\sigma
_{1})$ and $({\mathbf{k}}_2$, ${{E}_2}$, $\sigma _{2})$, we attach the factor 
\[
\quad \qquad -L^{-3} \, \nu(|{\mathbf{k}}_1-{\mathbf{k}}_1^{\prime}|) \,
 L({\mathbf{k}}_1, \, {\mathbf{k}}_1^{\prime}) \, L({\mathbf{k}}_2, \, {\mathbf{k}}_2^{\prime}).
\]
\noindent To a double vertex of type (b), with two outgoing line ends 
$({\mathbf{k}}_1^{\prime}$, ${E_1}^{\prime}$, $\sigma _{1})$%
and  $({\mathbf{k}}_2^{\prime}$, ${{E}_2}^{\prime}$, $\sigma _{2})$
along with two outgoing line ends $(-%
{\mathbf{k}}_1$, $-E_1$, $-\sigma _{1})$ and $(-{\mathbf{k}}_2$, $-{{E}_2}$, $-\sigma _{2})$, as well as to a double vertex of type (c) with two
incoming line ends with indices $(-{\mathbf{k}}_1^{\prime}$, $-{{%
E}_1}^{\prime}$, $-\sigma _{1})$ and $(-{\mathbf{k}}_2^{\prime}$, $-{{E}_2}^{\prime}$,  $-\sigma _{2})$ along with two incoming
line ends with indices ${\mathbf{k}}_1$, ${{%
E}_1}$, $\sigma _{1})$ and $({\mathbf{k}}_2$, ${{E}_2}$, $\sigma _{2})$, we attach the factor 
\[
 \quad - L^{-3}\, \nu(|{\mathbf{k}}_1 -{\mathbf{k}}_1^{\prime}|)\,4\sigma _{1}\sigma _{2}
 \, M({\mathbf{k}}_1, \, {\mathbf{k}}_1^{\prime}) \, M({\mathbf{k}}_2, \, {\mathbf{k}}_2^{\prime}).
\]

\begin{figure}[htbp]
\hspace{-0.5cm}\epsfig{figure=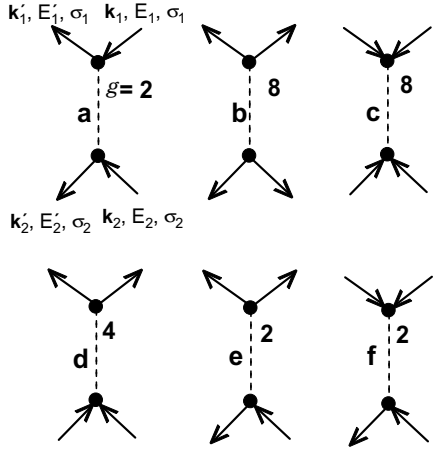,width=8.0cm}
\caption[]{Six different types of vertices with corresponding topological
automorphism factors g.}
\end{figure}

\smallskip
\noindent {\bf 3. Coupled BS equations}
\smallskip
\smallskip

Because of different kinds of vertices in Fig. 1 we now have a system of {\it two} coupled BS equations. Figure 2
shows their diagrammatic representation in the ladder approximation
of the two-electron BS equation for the BCS ground state. Depicted are both
the two-electron $\psi _{+}({\mathbf{k}}E;{\mathbf{K}}{\mathcal{%
E}}_K)$ and two-hole $\psi _{-}({\mathbf{k}}E; {\mathbf{K
}}{\mathcal{E}}_K)$ bound-state functions. Here ${\mathbf{K
}} \equiv {\mathbf{k}_1}+ {\mathbf{k}_2}$ is the total (or center-of-mass) wave-vector and ${\mathcal{E}}_K$ is the total energy of the two electrons referred to $2 E_F$.

Using the diagramatic rules just described, the two-component BS equations   for the bound state is
\[
 \psi _{+}({\bf k}E)=-\frac{1}{2\pi i}%
\int_{-\infty}^{+\infty}
dE^{^{\prime }}\,{L^{-3}}\,%
{\sum}_{{\bf k}^{^{\prime }}}%
\,\nu (|{\bf k}-{\bf k}^{\prime }|)\times
\]
\vspace{-0.55cm}
\[
 \times \,L({\bf K}/2+{\bf k},{\bf K}/2+{\bf k}^{\prime }) \, 
L({\bf K}/2-{\bf k}, {\bf K}/2-{\bf k}^{\prime })\times 
\]
\vspace{-0.69cm}
\[
\times \, ( i/\hbar)^{2}{\cal G}_{0}\left( {\bf K/}2+%
{\bf k},{\cal E}_K/2+E\right) \times 
\]
\vspace{-0.69cm}
\[
\times {\cal G}_{0}\left( {\bf K/}2-{\bf k},{\cal E}_K/2-E\right) \,\psi _{+}(%
{\bf k}^{^{\prime }}E^{^{\prime }})- 
\]
\vspace{-0.69cm}
\[
-\frac{1}{2\pi i}%
\int_{-\infty}^{+\infty}
dE^{^{\prime }}\,{L^{-3}} \,%
 {\sum}_{{\bf k}^{^{\prime }}}%
\, \nu (|{\bf k}-{\bf k}^{\prime}|)\times 
\]
\vspace{-0.55cm}
\[
 \times \,M({\bf K}/2+{\bf k}, {\bf K}/2+{\bf k}^{\prime }) \, 
M({\bf K}/2-{\bf k}, {\bf K}/2-{\bf k}^{\prime })\times 
\]
\vspace{-0.69cm}
\[
 \times \, (i/\hbar)^{2}{\cal G}_{0}\left( {\bf K/}2+%
{\bf k},{\cal E}_K/2+E\right) \times 
\]
\vspace{-0.8cm}
\begin{equation}
\times {\cal G}_{0}\left( {\bf K/}2-{\bf k},{\cal E}_K/2-E\right) \,\psi _{-}(%
{\bf k}^{^{\prime }}E^{^{\prime }}); 
\label{Go1}
\end{equation}

\[
 \psi_{-}({\bf k}E)=-\frac{1}{2\pi i}%
 \int_{-\infty }^{+\infty}
dE^{^{\prime }}\,{L^{-3}}\,%
{\sum}_{{\bf k}^{^{\prime }}}%
\, \nu (|{\bf k}-{\bf k}^{\prime}|)\times 
\]
\vspace{-0.55cm}
\[
\times \,L({\bf K}/2+{\bf k,K}/2+{\bf k}^{\prime }) \, 
L({\bf K}/2-{\bf k,K}/2-{\bf k}^{\prime })\times 
\]
\vspace{-0.69cm}
\[
 \times \,(i/\hbar)^{2}{\cal G}_{0}\left( -{\bf K/}2-%
{\bf k},-{\cal E}_K/2-E \right) \times 
\]
\vspace{-0.69cm}
\[
\times {\cal G}_{0}\left( -{\bf K}/2+{\bf k},-{\cal E}_K/2+E\right) \,\psi
_{-}({\bf k}^{^{\prime }}E^{^{\prime }})- 
\]
\vspace{-0.69cm}
\[
-\frac{1}{2\pi i}%
\int_{-\infty}^{+\infty}
dE^{^{\prime }}\,{L^{-3}} \,
{\sum}_{{\bf k}^{^{\prime }}}%
\, \nu (|{\bf k}-{\bf k}^{\prime}|)\times \, 
\]
\vspace{-0.55cm}
\[
 \times M({\bf K}/2+{\bf k,K}/2+{\bf k}^{\prime }) \, M({\bf K}/2-{\bf k,K}/2-{\bf k}^{\prime })\times 
\]
\vspace{-0.69cm}
\[
 \times \, (i/\hbar)^{2}{\cal G}_{0}\left( -{\bf K/}2-%
{\bf k},-{\cal E}_K/2-E\right) \times 
\]
\vspace{-0.69cm}
\begin{equation}
 \times {\cal G}_{0}\left( -{\bf K/}2+{\bf k},{\cal E}_K/2+E\right) \,\psi _{+}(%
{\bf k}^{^{\prime }}E^{^{\prime }}).
\label{Go2}
\end{equation}

\noindent It can be shown that these equations coincide exactly with those used in the description
of collective excitations in the BCS-Bogoliubov\noindent \cite{BTS}
\noindent microscopic theory of superconductivity. \\

\begin{figure}[t]
\hspace{-0.5cm}\epsfig{figure=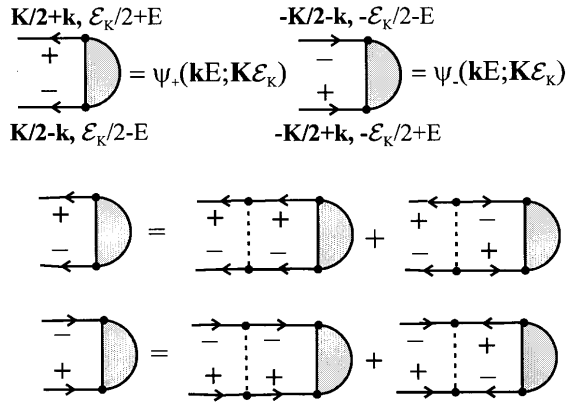,width=8.0cm}
\caption[]{Diagrammatic representation of the two-com- ponent coupled BS equations.}
\label{Fig 2}
\end{figure}

\smallskip
\noindent {\bf 4. CP solutions of BS equations}
\smallskip
\smallskip

We now employ the BCS model interaction 
\begin{equation}
 \nu (|{\mathbf{k}}-{\mathbf{k}}^{\prime}|)\Rightarrow -(k_{F}^{2}/{k^{\prime}}^2) \, V \, \eta(k) \eta(k^{\prime}),\,   
 \label{interaction}
\end{equation}

\noindent where $V \geq 0$, and  $\eta(k) = 1$ when $k_{F}-k_D
< k < k_{F}+ k_D$ and $= 0$ otherwise. Here $k_D \equiv m \omega_D/ \hbar k_F$ with $\omega_D$ the Debye frequency, if $\hbar \omega_D << E_F$. 

The detailed solution for the system of the two-component coupled BS equations of Fig. 2 is too
cumbersome to present here, but it can be shown that they yield {\it two
types} of independent solutions. \noindent The {\it first} solution of (\ref{Go1}) and (\ref{Go2}) with (\ref{interaction}) is just 
\[
{{V \, k_{F}^{2}} \over {2\pi ^{2}}} \int_{k_{F}- k_D}^{
k_{F}+  k_D} dk \int_{-1}^{1} dt\,\,u(\left| 
{\mathbf{K}}/2 + {\mathbf{k}} \right| )\, \times
\]
\vspace{-0.5cm}
\[
\times v(\left| {\mathbf{K}}/2 - {\mathbf{k}} \right|) \left[u( \left| {\mathbf{K}}/2-{\mathbf{k}} \right| ) \,v( \left| {\mathbf{K}}/2 +
{\mathbf{k}} \right|)- \right.
\]
\vspace{-0.7cm}
\[ 
 - \left. u(\left| {\mathbf{K}}/2+{\mathbf{k}}\right| )\,v(\left| 
{\mathbf{K}}/2-{\mathbf{k}} \right| ) \right] \times 
\]
\vspace{-0.7cm}
\begin{equation}
 \frac{E (|{\mathbf{K}}/2+ {\mathbf{k}}|)+ E(| {\mathbf{K}}/2- {\mathbf{k}}|) }{ -
{\mathcal{E}}_K^{2} + [E(|{\mathbf{K}}/2 + {\mathbf{k}}|) + E(|{\mathbf{K}}/2-
{\mathbf{k}}|) ]^{2} } = 1. 
\label{uv-uv}
\end{equation}

\noindent Here $t \equiv \cos \theta $,  $\theta$ being the angle between $%
{\mathbf{k}}$ and ${\mathbf{K}}$ while $\left| {\mathbf{K}}/2+%
{\mathbf{k}}\right| =\sqrt{k^{2}+K\,k\,t+K^{2}/4}$ and  $\left| {\mathbf{K}}/2-%
{\mathbf{k}}\right| =\sqrt{k^{2}-K\,k\,t+K^{2}/4}.$ 

As we see the numerator in (\ref{uv-uv}) vanishes for $K=0$, so the denominator must vanish as $K \rightarrow 0$ as well. This gives ${\mathcal{E}}_K = \pm 2\,\Delta + O(K).$ To find the asymptotic solution for both small
coupling and $K$ we may neglect terms with $\Delta $ in $u,v$ in the
integrand. Thus we put $u(k) \simeq \theta
(k_{F}-k)\equiv \theta _{F}(k)$, $v(k)\simeq \theta (k-k_{F})\equiv \theta _{G}(k),
$ \noindent where $\theta (k)$ is the usual Heaviside step function. Therefore,
\begin{eqnarray*}
& u(\left| {\mathbf{K}}/2+{\mathbf{k}}\right| )\,v(\left| {\mathbf{K
}}/2-{\mathbf{k}}%
\right| )\times & \\
 &  \qquad \times \lbrack u(\left| {\mathbf{K}}/2-{\mathbf{k}}\right|
)\,v(\left| {\mathbf{K}}/2+{\mathbf{k}}\right| ) - & \\
  & -u(\left| {\mathbf{K}}/2+{\mathbf{k}}%
\right| )\,v(\left| {\mathbf{K}}/2-{\mathbf{k}}\right| )] \simeq & \\
\\
  & -\theta _{F}(\left| {\mathbf{K}}/2+{\mathbf{k}}\right| )\,\theta _{G}(\left| 
{\mathbf{K}}/2-{\mathbf{k}}\right| ). & 
\end{eqnarray*}

\noindent\ Thus we can rewrite (\ref{uv-uv}) as
\[
  \frac{V\,k_{F}^{2}}{4\pi ^{2}} \int_{k_{F}-k_{D}}^{k_{F}+k_{D}} dk \int_{0}^{1} dt \, \times  
 \]
\vspace{-0.4cm}
\[
\times \, \theta _{F}\left( \left| {\mathbf{K}}/2+
{\mathbf{k}}\right| \right) \theta_{G} \left( \left| {\mathbf{K}}/2-%
{\mathbf{k}}\right|\right) \times 
 \]
\vspace{-0.4cm}
\[
 \times \, \left[  \frac{1}{-{\mathcal{E}}_K+E\left( \left| {\mathbf{K}}/2+%
{\mathbf{k}}\right| \right) + E\left( \left| {\mathbf{K}}/2-%
{\mathbf{k}}\right| \right) } \right. + \quad 
\]
\vspace{-0.4cm}
\begin{equation}
\quad \left.  +\frac{1}{{\mathcal{E}}_K+E \left( \left| {\mathbf{K}}/2+%
{\mathbf{k}}\right| \right) + E\left( \left| {\mathbf{K}}/2-%
{\mathbf{k}}\right| \right) } \right] = 1. 
\label{energy}
\end{equation}

\noindent In Fig. 3 we show as shaded the region of integration over $k$ and $t=\cos \,\theta $. The integral over the polar angle $%
\theta $ is restricted to from $0$ to $\pi /2$ since the integrand in (\ref{uv-uv}) vanishes for $t<0$. We first integrate over $k$ from $%
k_{\min }$ to $k_{\max }$ which are solutions of 
$\left| {\mathbf{K}}/2+{\mathbf{k}}\right| =k_{F}$, and  $|%
{\mathbf{K}}/2-{\mathbf{k|}}=k_{F},$ \noindent or \noindent\ 
\begin{eqnarray*}
k_{\min } & = & -Kt/2+\sqrt{K^{2}t^{2}/4-{K^{2}}/{4}+k_{F}^{2}}, \\
k_{\max } & = &Kt/2+\sqrt{K^{2}t^{2}/4-{K^{2}}/{4}+k_{F}^{2}}.
\end{eqnarray*}

\noindent The restrictions $k_{F}-k_{D}\leq k\leq k_{F}+k_{D}$ are not
illustrated in Fig. 3 as they are unimportant for us. \noindent As $K\rightarrow 0$ we then have 
\begin{eqnarray*}
 & k_{\min } \simeq -Kt/2+O(K^{2}), & \\
& k_{\max } \simeq Kt/2+O(K^{2}), & \\
& E\left( \left| {\mathbf{K}}/2+%
{\mathbf{k}}\right|\right) +E\left( \left| {\mathbf{K}}/2-%
{\mathbf{k}}\right|\right) \simeq 2\,\Delta +O(K^{2}) &
\end{eqnarray*}

\noindent where we redefined the integration variable $k=k_F$ $+xK$, with $x$ 
 the new variable.

\begin{figure}[htbp]
\epsfig{figure=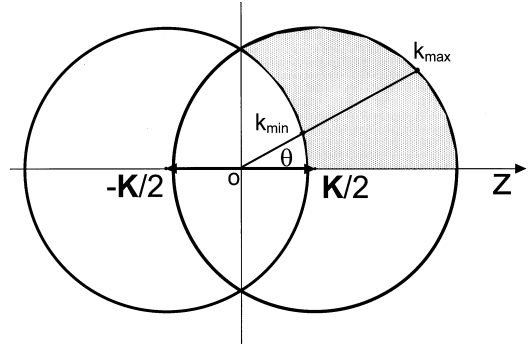,width=7.0cm}
\caption[]{Integration region in (\ref{energy})}
\label{Fig 3}
\end{figure}

\noindent Consequently in seeking a solution of Eq. (\ref{energy}), for example of the form 
\[
{\mathcal{E}}_K = 2\,\Delta +c \hbar K + O({K}^{2}),
\]

\noindent where $c$ is a constant, we obtain for small $\Delta $ and 
${K}$ 
\[
\frac{V\,k_{F}^{2}}{4\pi ^{2}\hbar } \int_{0}^{1} dt%
\int_{k_{F}- K t/2}^{k_{F}+ K t/2}  \frac{{\rm d}k}{c K } \simeq 1.
\]

\noindent Thus $c = V\,k_{F}^{2}/8\pi ^{2} \hbar $.
Finally one gets

\begin{equation}
{\mathcal{E}}_K = 2\Delta + \frac{\lambda}{4} \,v_{F}\,\hbar K + O(K^2), 
\label{energ+}
\end{equation}

\noindent together with its symmetric solution 
\begin{equation}
{\mathcal{E}}_K=-2\Delta - \frac{\lambda}{4} \,v_{F}\,\hbar K + O(K^2),
\label{energ-}
\end{equation}
where $\lambda \equiv N(0)V$ and $N(0)\equiv mk_F/2 \pi^2 \hbar^2$. 
 These two solutions describe {\it moving} $2p$-CPs and $2h$-CPs,
respectively. A linear -in-$K$ behavior of the moving pair binding energy was
obtained for the original Cooper-pair problem \cite{Llan} but it was independent of the interaction coupling strength as it excluded $2h$-CP contributions. More significantly, the binding energy there was {\it negative} as it refers to an infinite-lifetime composite particle, while in  (\ref{energ+}) it is {\it positive} as it describes a resonance in the continuum with a finite lifetime as evidenced by an imaginary contribution \cite{tolma2} appearing in higher order terms in $K$. \\

\smallskip
\noindent {\bf 5. ABH-like mode solution of BS equations}
\smallskip
\smallskip

The {\it second} solution of the coupled BS equation (\ref{Go1}) and (\ref{Go2}) follows from
\[
 \frac{V\,k_{F}^{2}}{2\pi ^{2}}
\int_{k_{F}-k_{D}}^{k_{F}+k_{D}} dk \int_{-1}^{1} dt\,v(\left| {\mathbf{K}}/2+{\mathbf{k%
}}\right|) \times  
\]
\vspace{-0.5cm}
\[
 \times v(\left| {\mathbf{K}}/2-{\mathbf{k}}\right| )\lbrack u(\left| {\mathbf{K}}/2+ {\mathbf{k}}\right| )\,u(\left| {\mathbf{K}}/2-%
{\mathbf{k}}\right| )+ 
\]
\vspace{-0.5cm}
\[
 +v(\left| {\mathbf{K}}/2+{\mathbf{k}}\right| )\,v(\left| {\mathbf{K}}/2-{\mathbf{k}}%
\right| )] \times 
\]
\vspace{-0.5cm}
\[
 \times \frac{1}{{\mathcal{E}}_K+E(|{\mathbf{K/}}2+{\mathbf{k|)}}+E(|{\mathbf{K/}}2-%
{\mathbf{k|)}}}+ 
\]
\vspace{-0.5cm}
\[
 + \frac{V\,k_{F}^{2}}{2\pi ^{2}}
\int_{k_{F}-k_{D}}^{k_{F}+k_{D}} dk \int_{-1}^{1} dt \,u(\left| {\mathbf{K}}/2+{\mathbf{k%
}}\right| )\,\times 
\]
\vspace{-0.5cm}
\[
 \times u(\left| {\mathbf{K}}/2-{\mathbf{k}}\right| ) \lbrack u(\left| {\mathbf{K}}/2+{\mathbf{k}}\right| )\,u(\left| {\mathbf{K}}/2-%
{\mathbf{k}}\right| )+ 
\]
\vspace{-0.5cm}
\[
+v(\left| {\mathbf{K}}/2+{\mathbf{k}}\right| )\,v(\left| {\mathbf{K}}/2-{\mathbf{k}}%
\right| )] \times
\]
\vspace{-0.5cm}
\begin{equation}
 \times \frac{1}{-{\mathcal{E}}_K+E(|{\mathbf{K/}}2+{\mathbf{k|)}}+E(|{\mathbf{K/}}2-%
{\mathbf{k|)}}}=1. 
\label{energy2}
\end{equation}  

\noindent Proceeding as before, in the limit $K\rightarrow 0$ one now finds
\begin{equation}
{\mathcal{E}}_K= (v_{F}\,\hbar K /\sqrt{3})[1+\frac{\hbar \omega_D}{4E_F}e^{-2/\lambda}+ \cdots ] + O(K^2),
\end{equation}
which is similar to the Anderson-Bogoliubov-Higgs (ABH) RPA excitation mode in the BCS theory of superconductivity \cite{BTS,And,Higgs,Higgs2}. \\ 

\smallskip
\noindent {\bf 6. Conclusion}
\smallskip
\smallskip

We have presented a new many-fermion formalism based on a BS equation
applied to the full BCS ground state which does {\it not} neglect the presence of holes. This leads in the ladder approximation to two
types of solutions: the first referring to simple moving CPs consisting of  two-electron (12) {\it or} two-hole (13) resonances. In addition, this formalism naturally provides a second solution (15) of an entirely different physical nature which is analogous to the
ABH excitation mode. Bose-Einstein condensation can occur with the first type of objects but not with the second.

\vspace{0.5cm}

\noindent {{\bf Acknowledgments.}} Partial support from UNAM-DGAPA-PAPIIT (Mexico) \# IN102198 and CONACyT (Mexico) \# 27828 E is gratefully acknowledged.
V.V.T. acknowledges a CONACyT chair fellowship at UNAM.
\vspace{0.5cm}

*{On leave from N.E. Baumann State Technical University, 107005, 2-ja Baumanscaja Street, 5, Moscow, Russia}.

\end{document}